\documentstyle[12pt]{article}
\newcommand{\be}{\begin{equation}}
\newcommand{\ee}{\end{equation}}
\newcommand{\bea}{\begin{eqnarray}}
\newcommand{\eea}{\end{eqnarray}}
\newcommand{\s}{\sin}
\newcommand{\k}{\cos}
\newfont{\bg}{cmr10 scaled\magstep4}

\newcommand{\bigzerou}{\smash{\lower1.7ex\hbox{\bg 0}}}

\begin{document}
\baselineskip=24pt
\title{
\begin{flushright}
{\normalsize Gunma-Tech-95-1}
\end{flushright}
\vspace{20mm}
Instanton effects and Witten complex in supersymmetric quantum mechanics on
$SO(4)$}
\author{Kazuto Oshima \\ \\
\sl Gunma College of Technology,Maebashi 371,Japan}
\date{ }
\maketitle
\vspace{20mm}
We examine supersymmetric quantum mechanics on $SO(4)$ to
realize Witten's idea. We find instanton solutions connecting
approximate vacuums. We calculate Hessian matrices for these
solutions to determine true vacuums.
Our result is in agreement with de Rham cohomology of $SO(4)$.
We also give a criterion for cancellation of instanton effects for a pair of
instanton paths.
\vspace{20mm}
\\
\newpage
\section{Introduction}
In the last two decades theoretical
physics and mathematics have been greatly  developed through mutual stimulation
[1]. At this present Seiberg-Witten theory[2] has attracted the attention of
both
theoretical physicists and mathematicians.
In this paper we discuss Witten's pioneering work on supersymmetric
quantum mechanics and de Rham theory proposed in 1982[3].  He has given a
quantum mechanical
interpretation about de Rham theory on a Riemannian manifold $M$. He has
considered certain supersymmetric quantum mechanics on the manifold.
His idea is to adopt a superpotential which is obtained from
the Morse function $h$ of the manifold.
For each critical point of $h$, an approximate vacuum
can be identified. By instanton effects some of them found to be false
vacuums. True vacuums of the theory correspond to harmonic forms on the
manifold.  Witten's conjecture is that the number of true vacuums
agrees with the dimension of de Rham cohomology. The diminishing of the
number of vacuums implies the Morse inequality.

His idea has been established in the operator formalism[4]
However, in the Laglangian formalism the Witten's program has been performed
only for few cases. The case $M={\bf R}$ has well been investigated
in the context of supersymmtry breaking[5]. Recently, Yasui et al.[6]
have investigated the case $M=SO(3)$. It has been shown that among four
approximate vacuums two survive as true vacuums under instanton effects
in agreement with de Rham cohomology of $M=SO(3)$.
Our purpose in this paper is to
examine a more complicated case $M=SO(4)$ in the Laglangian formalism.
We discuss a condition for occurrence of instanton effects and identify
true vacuums.

In Sec.2 supersymmetric quantum mechanics on a manifold $M$ is formulated.
In Sec.3 $M=SO(2)$ case is discussed for later convenience. In Sec.4
coordinates,metric and critical points for $M=SO(4)$ are treated.
In Sec.5 Hessian matrices are calculated and instanton effects are discussed.
Section 6 is devoted to summary
and discussion.
\section{Supersymmetric Quantum Mechanics on a Manifold}
In this section the relevant supersymmetric quantum mechanics on a
manifold $M$ is formulated. In particular, approximate vacuums, a gradient flow
equation and Hessian matrices are described.
The supersymmetric quantum mechanics on $M$ is derived from the following
Laplacian:
\begin{equation}
\hat{H}={1 \over 2}(d_{h} d_{h}^{\dagger} +d_{h}^{\dagger} d_{h}),
\end{equation}
where $d_{h}=e^{-h}de^{h}, d_{h}^{\dagger}=e^{h}d^{\dagger}e^{-h},
d$ is the exterior derivative and $d^{\dagger}$ is its adjoint operator.
Fermion creation and annihilation operators $\hat{\psi} ^{*}{}^{i}$ and
$\hat{\psi} ^{i}$
can be identified with the exterior multiplication $e_{dx ^{i}}$ and the
interior multiplication $i_{\partial \over \partial x^{i}}$. Subsequently,
on a flat metric
$d=\hat{\psi} ^{*}{}^{i} {\partial \over \partial x^{i}}, d^{\dagger}=
\hat{\psi} ^{i}{\partial \over \partial x^{i}}$
and $\hat{\psi} ^{*}{}^{i_{1}} {\ldots}
\hat{\psi} ^{*}{}^{i_{m}}|0>_{F}$ can be identified with bases of $m$-forms[3].
An approximate vacuum $|\Omega _{l}>$ localized near a critical point
$P^{(l)}$ satisfies
\begin{equation}
d_{h}|\Omega _{l}>=d_{h}^{\dagger}|\Omega _{l}>=0
\end{equation}
up to quantum effects. Some of approximate vacuums cease to satisfy
(2.2) by quantum effects. If $|\Omega _{l}>$ is a true vacuum, it remains
satisfying (2.2) after taking quantum effects into account. Because
$d_{h}^{\dagger}$ is the adjoint of $d_{h}$ , states satisfying (2.2)
are elements
in $\displaystyle{ Ker d_{h} \over Im d_{h}}$ called the Witten
complex. We examine $d_{h}|\Omega _{l}>$
quantum mechanically on $M=SO(4)$.
If $<\Omega _{l+1}|d_{h}|\Omega _{l}>
\neq 0$, neither approximate vacuum is a true vacuum. According to [3],
the following form is valid:
\begin{equation}
<\Omega _{l+1}|d_{h}|\Omega _{l}>= \sum_{\gamma}
n_{\gamma} e^{-(h(P^{(l+1)})-h(P^{(l)}))},
\end{equation}
where $n_{\gamma}$ is an integer assigned for each instanton path
$\gamma$ .

The supersymmetric hamiltonian derived from (2.1) is
\begin{equation}
2\hat{H} =-g^{-{1 \over 2}} \nabla _{i}g^{1 \over 2}g^{ij} \nabla _{j}
+R_{ijkl}\hat{\psi}^{k}\hat{\psi} ^{*}{}^{l}\hat{\psi}^{*}{}^{j}\hat{\psi} ^{i}
+g^{ij} {\partial h \over \partial x^{i}}{\partial h \over \partial x^{j}}
+H_{ij}\{\hat{\psi} ^{*}{}^{j},\hat{\psi}^{*}{}^{i}\},
\end{equation}
where $g_{ij}$ and $R_{ijkl}$ are the Riemann metric and tensor, and
$\nabla _{i}=
{\partial \over \partial x^{i}}-\Gamma_{ik}^{l}\hat{\psi} ^{*}{}^{k}
\hat{\psi}_{l}$ is the covariant derivative; $H_{ij}$ is the Hessian matrix
\begin{equation}
H_{ij}=(\partial_{i}\partial_{j}-\Gamma_{ij}^{k}\partial_{k})h.
\end{equation}
Corresponding Laglangian is
\begin{equation}
{\cal L} ={1 \over 2}g_{ij}{dx^{i} \over dt}{dx^{j} \over dt}
+{1 \over 2}g^{ij}{\partial h \over \partial x^{i}}
{\partial h \over \partial x^{j}}
+\psi ^{*}{}^{i}({d \over dt}\psi_{i}-\Gamma_{ij}^{k}\psi_{k}
{dx^{j} \over dt})+H_{ij}\psi ^{*}{}^{j}\psi^{i}
+{1\over4}R_{ijkl}\psi^{i}\psi^{j}\psi^{*}{}^{k}\psi^{*}{}^{l}.
\end{equation}

Classical solutions give main contribution to a pathintegral.
In a supersymmetric model, we only have to consider quasi classical
solutions satisfying equation of motion with fermion disregarded.
Quasi classical solutions of $(2.6)$ obey
\begin{equation}
{dx^{i} \over dt}=\pm g^{ij}{\partial h \over \partial x^{j}}.
\end{equation}
By the operation of $d_{h}$, the Morse index $l$ identical to the
number of excited fermions increases by one. We choose one of the
sign in $(2.7)$ so that the value of the Morse function $h$ increases.

For $M=SO(3)$, Eq.(2.7) has a pair of
instanton solutions connecting two critical
points whose Morse indices differ by one. Corresponding action for
each instanton solution is written to the one-loop level as[6]
\begin{equation}
S=S_{cl}+ \int _{-\infty}^{\infty}
 dt\{-{1 \over 2}\xi^{a}({d \over dt}+\lambda _{a}(t))
({d \over dt}-\lambda _{a}(t))\xi^{a}+\psi ^{a}({d \over dt}-\lambda _{a}(t))
\psi _{a}^{*}\},
\end{equation}
where $\xi^{a}$ are some linear combinations of geodesic coordinates around
the instanton path and $t$ is the time parameter of the instanton mode;
$\lambda _{a}(t)$ are eigenvalues of the Hessian matrix
$H_{i}^{j}=H_{ik}g^{kj}$
at a time $t$. Corresponding approximate hamiltonian is
\be
2\tilde{H}=\sum _{a}(-({\partial  \over \partial \xi ^{a}})^{2}
+\lambda _{a}^{2}(t)(\xi ^{a})^{2}+\lambda _{a}(t)\{\hat{\psi}_{a}^{*},
\hat{\psi}^{a}\}).
\ee
An approximate vacuum near a critical point with the Morse index $l$ is
expressed as
\be
|\Omega _{l}>=\pi ^{-{3 \over 2}} \prod_{a} |\lambda_{a}|^{1 \over 4}
\rm{exp}(-{1 \over 2} \sum_{a} |\lambda_{a}|(\xi ^{a})^{2})
\hat{\psi} _{b _{1}}^{*}\ldots \hat{\psi} _{b _{l}}^{*}|0>_{F},
\ee
where $b_{1},\ldots ,b_{l}$ correspond to negative eigenvalues
of the Hessian matrix $H_{i}^{j}$ at the critical point.
Like this, approximate vacuums are decided by eigenvalues of $H_{i}^{j}$.

Similar situation will hold for $M=SO(4)$. We calculate the Hessian
matrix $H_{i}^{j}$ to examine approximate vacuums. Moreover, we discuss
instanton effects between adjacent
approximate vacuums. Transition
amplitudes are calculated from eigenvalues of
$H_{i}^{j}$ up to signs[6].
Owing to the notorious minus signs associated with fermions, it is not
easy to determine the signs. It is crucial to determine the signs,
because a pair of instanton effects can cancel each other.
\section{$SO(2)$ Case}
\setcounter{equation}{0}

In this section we discuss the case $SO(2)$.
We see how instanton effects occur or disappear.

We denote a group element in $SO(n)$ by $A=(a_{ij})$. The Morse
function $h$ for $SO(n)$ is given by[7]
\be
h=\sum _{a} c_{i}a_{ii} , (c_{i}>2c_{i+1}>0).
\ee
For $M=SO(2)$, $A$ is represented as
\begin{eqnarray}
    A=\left(
     \begin{array}{cc}
       \cos \theta & -\sin \theta \\
       \sin \theta & \cos \theta
     \end{array}
     \right) .
\eea
The Morse function is $h=(c_{1}+c_{2}) \cos \theta$ and the critical points
are
\be
P^{(0)}=\rm{diag}(-1-1),    P^{(1)}=\rm{diag}(\quad 1\quad 1),
\ee
with the Morse indices $l=0$ and $l=1$ respectively.
Since $g_{\theta \theta}=g^{\theta \theta}=1$, the gradient flow
equation (2.7) simplifies to
\be
{d\theta \over dt}=-(c_{1}+c_{2})\sin \theta.
\ee
This equation has the following instanton solution
\be
\cos\theta=\tanh((c_{1}+c_{2})t+\alpha),
\ee
with an arbitrary constant $\alpha$. Corresponding to $\sin \theta \ge 0$
or  $\sin \theta \le 0$, there are two paths connecting the critical points.
For the latter path, we introduce a new coordinate $\theta ^\prime$
by $\theta ^\prime=-\theta$. We call the coordinates $\theta$ and
$\theta ^\prime$ odd, because they are opposite in the sign. Thus, we
only have to consider the path $0 \le \theta \le \pi$ in each coordinate
system. We have now two instanton paths
\begin{eqnarray}
     \left(
     \begin{array}{cc}
       \cos \theta & \mp \sin \theta \\
       \pm \sin \theta & \cos \theta
     \end{array}
     \right) \hspace{1cm} ( \sin\theta \ge 0)
\end{eqnarray}
with $\cos \theta$ (3.5).

These instanton paths lead to transitions between approximate vacuums
\bea
&|0>& \longrightarrow |1> \equiv \hat{\psi ^{*}}|0> \equiv |\theta > \sim
d\theta |0>
,\nonumber \\
&|0>& \longrightarrow \hat{\psi^{\prime}{}^{*}}|0> \equiv |\theta ^{\prime}>
 \sim d\theta ^{\prime} |0>
,
\eea
where the state $|0>$ on the left(right) hand side means
a bosonic(bosonic parts of an)
approximate vacuum around the critical point $P^{(0)}(P^{(1)})$;
the state $|1>$ implies one fermionic mode is excited;moreover, we mean
by $|\theta>$ a fermionic mode corresponding to the coordinate $\theta$
is excited;the symbol $\sim$ means the identification of a $p$ fermion
excited state and a $p-$form[3].
Since $d\theta ^{\prime} = -d\theta$, we see $\hat{\psi} ^{\prime}{}^{*}|0>
=-\hat{\psi} ^{*}|0>$. The appearance of the relative minus sign for a
pair of fermionic excitations is a general feature for an instanton mode.
By the operator $d_{h}$, the approximate vacuum $|0>$ transforms as
\be
d_{h}|0>= e^{-2(c_{1}+c_{2})} (\hat{\psi}^{*}+\hat{\psi^{\prime *}})
|0>=0,
\ee
and the matrix element $<1|d_{h}|0>$ vanishes. Thus, there is no instanton
effect between the two approximate vacuums $|0>$ and $|1>$. Both of them
remain to be true vacuums. We have one true vacuum at each critical point
in agreement with the result
of de Rham cohomology; $H^{0}(SO(2))=H^{1}(SO(2))
=\bf R $.

As we have seen, a pair of instanton paths lead to fermionic states
with opposite signs and there is no instanton effect. In the present case
the original state is $|0>$,  which is common to both instanton paths.
In general, this is not the case. An approximate vacuum may provide
opposite signs for a pair of instanton paths.

We call a state,like $|0>$, common to a pair of instanton paths
an even state. We call a state, like $\hat{\psi} ^{*} |0>$, providing
different signs for a pair of instanton paths  an odd state.
We call the evenness and oddness parity. We can summarize the condition
instanton effects occur. An approximate vacuum has definite parity for
a pair of instanton paths. Instanton effects do not cancel
each other for a pair of
paths only when the parity of an approximate vacuum  does not change along
the paths. This judgment will be used later.
\section{Coordinates, Metric and Critical Points on $SO(4)$}
\setcounter{equation}{0}
In this section we first introduce a coordinate system and an invariant
metric on $M=SO(4)$. Let us introduce the generalized Euler angles[8].
Consider the following three infinitesimal generators of $SO(4)$:
\bea
  [E_{1} = \left( \begin{array}{cccc}
            0&1&0&0\\
            -1&0&0&0\\
            0&0&0&0\\
            0&0&0&0\\
            \end{array} \right)  ] ,
  [E_{2} = \left( \begin{array}{cccc}
            0&0&0&0\\
            0&0&1&0\\
            0&-1&0&0\\
            0&0&0&0\\
            \end{array} \right)  ],
 [E_{3} = \left( \begin{array}{cccc}
            0&0&0&0\\
            0&0&0&0\\
            0&0&0&1\\
            0&0&-1&0\\
            \end{array} \right)  ].
\eea
A group element $A$ in $SO(4)$ can be parametrized as
\bea
{\it A} = e^{cE_{1}}e^{zE_{2}}e^{yE_{3}}e^{bE_{1}}e^{xE_{2}}e^{aE_{1}}=
\nonumber  \\
\vspace{10mm}
\begin{tiny}
 \left( \begin{array}{llll}
    (\cos b \cos c - \sin b \sin c \cos z)\cos a &
    (\cos b \cos c - \sin b \sin c \cos z)\sin a &
   (\sin b \cos c + \cos b \sin c \cos z)\sin x  &
    \sin c \sin y \sin z \cr

   +(-(\sin b \cos c + \cos b \sin c \cos z) \cos x &
   +((\sin b \cos c + \cos b \sin c \cos x) \cos x &
   +\sin c \cos x \cos y \sin z & \cr
\vspace{5 mm}
   +\sin c \sin x \cos y \sin z)\sin a  &
   - \sin c \sin x \cos y \sin z)\cos a & & \cr

   -(\cos b \sin c + \sin b \cos c \cos z)\cos a &
   -(\cos b \sin c + \sin b \cos c \cos z)\sin a &
   (- \sin b \sin c + \cos b \cos c \cos z) \sin x &
    \cos c \sin y \sin z \cr

   +((\sin b \sin c - \cos b \cos c \cos z) \cos x &
   -((\sin b \sin c - \cos b \cos c \cos z) \cos x &
   +\cos c \cos x \cos y \sin z & \cr
\vspace{5 mm}
   + \cos c \sin x \cos y \sin z)\sin a &
   +\cos c \sin x \cos y \sin z)\cos a & & \cr

   \sin x( \cos y \cos z + \cos b \cos x \sin z)\sin a  &
    (\sin x \cos y \cos z + \cos b \cos x \sin z)\cos a  &
    \cos x \cos y \cos z - \cos b \sin x \sin z &
    \sin y \cos z \cr
\vspace{5 mm}
   +\cos a \sin b \sin z  &
 + \sin a \sin b \sin z  & & \cr

   -\sin a \sin x \sin y, & \cos a \sin x \sin y &
   -\cos x \sin y & \cos y \cr
\end{array}
    \right)
\end{tiny}
\nonumber
\eea
\be
\ee
In appendix A left invariant vector fields are noted.
{}From (A.4)
an $SO(4)$-invariant metric is obtained:
\bea
(\sin^{2} y \ \ g^{ij})=  \\
\begin{tiny}
\bordermatrix{
\vspace{5 mm} & a & b & c & x & z & y \cr
\vspace{5 mm}
a & \displaystyle{\frac1{\sin^{2} x}}
  & - \displaystyle{\frac {\cos x}{\sin^{2} x}}
    - \displaystyle{\frac {\cos b \cos y \cos z}{\sin x \sin z}}
  & \displaystyle{\frac{\cos b \cos y}{\sin x \sin z}}
  & 0 & - \displaystyle{\frac{\sin b \cos y}{\sin x}} & 0 \cr
b & - \displaystyle{\frac{\cos x}{\sin^{2} x}}
    - \displaystyle{\frac{\cos b \cos y \cos z}{\sin x \sin z}}
  & \cot ^{2} x
    + \cot ^{2} z
    + \sin^{2} y & - \displaystyle{\frac{\cos b \cos x \cos y}{\sin x \sin z}}
    - \displaystyle{\frac{\cos z}{\sin^{2} z}}
  & \displaystyle{\frac{\sin b \cos y \cos z}{\sin z}}
  & \displaystyle{\frac{\sin b \cos x \cos y}{\sin x}} & 0 \cr
\vspace{5 mm}
  & & + 2 \displaystyle{\frac{\cos b \cos x \cos y \cos z}{\sin x \sin z}}
  & & & & \cr
\vspace{5 mm}
c & \displaystyle{\frac{\cos b \cos y}{\sin x \sin z}}
  & - \displaystyle{\frac{\cos b \cos x \cos y}{\sin x \sin z}}
    - \displaystyle{\frac{\cos z}{\sin^{2} z}}
  & \displaystyle{\frac1{\sin^{2} z}} &
    - \displaystyle{\frac{\sin b \cos y}{\sin z}} & 0 & 0 \cr
\vspace{5 mm}
x & 0 & \displaystyle{\frac{\sin b \cos y \cos z}{\sin z}}
  & - \displaystyle{\frac{\sin b \cos y}{\sin z}}
  & 1 & - \cos b \cos y & 0 \cr
\vspace{5 mm}
z & - \displaystyle{\frac{\sin b \cos y}{\sin x}}
  & \displaystyle{\frac{\sin b \cos x \cos y}{\sin x}}
  & 0 & - \cos b \cos y & 1 & 0 \cr
\vspace{5 mm}
y & 0 & 0 & 0 & 0 & 0 & \sin^{2} y \cr
  }  ,
\end{tiny}
\eea
\begin{eqnarray*}
 (g_{ij}) =   \\
\begin{small}
 \bordermatrix{
\vspace{5 mm} & a & b & c & x & z & y \cr
    & & & \cos x \cos z & & \cr
\vspace{5 mm}
a & 1 & \cos x & - \cos b \sin x \cos y \sin z &
       0 & \sin b \sin x \cos y & 0 \cr
\vspace{5 mm}
b & \cos x & 1 & \cos z & 0 & 0 & 0 \cr
    & \cos x \cos z & & & & \cr
\vspace{5 mm}
c & - \cos b \sin x \cos y \sin z & \cos z &
       1 & \sin b \cos y \sin z & 0 & 0 \cr
\vspace{5 mm}
x &  0 & 0 & \sin b \cos y \sin z & 1 & \cos b \cos y & 0
\cr \vspace{5 mm}
z &  \sin b \sin x \cos y & 0 & 0 & \cos b \cos y & 1 & 0 \cr
\vspace{5 mm}
y &  0 & 0 & 0 & 0 & 0 & 1
  }
\end{small}
\end{eqnarray*}
\be
\ee
In appendix B non-zero Christoffel symbols for this metric are noted.

{}From (3.1) the Morse function $h$ is given by
\bea
  h=c_{1}(\k a \k b \k c-\k a \s b \s c \k z-\s a \s b \k c \k x
  \nonumber      \\+\s a \s c \s x \k y \s z-\s a \k b \s c \k x \k z)
  \nonumber \\
+ c_{2}(-\s a \k b \s c-\s a \s b \k c \k z -\k a \s b \s c \k x \nonumber \\
       -\k a \k a \s x \k y \s z + \k a \k b \k c \k x \k z) \nonumber \\
+c_{3}(\k x \k y \k z-\k b \s x \s z)+c_{4}\k y, \nonumber \\
(c_{1} > 2c_{2} > 4c_{3} >8c_{4}>0).
\eea
At critical points on $M$ the Morse function $h$ takes extremal values.
The Morse function $h$ (4.5) has eight critical points, which correspond
to diagonal rotations in $SO(4)$;
\bea
P^{(0)}&=&(-1 -1 -1 -1), P^{(1)}=(-1 -1 \quad 1 \quad 1),
P^{(2)}=(-1 \quad 1 -1 \quad 1),\nonumber \\
P^{(3A)}&=&(-1 \quad 1 \quad 1 -1),P^{(3B)}=(1 -1 -1 \quad 1),
P^{(4)}=(1 -1 \quad 1 -1), \nonumber \\
P^{(5)}&=&(1 \quad 1 -1 -1),P^{(6)}=(1 \quad 1 \quad 1 \quad 1).
\eea
In Fig.1 instanton paths connecting these critical points are noted.
\vspace{10mm}
\\
\vspace{10mm}
\begin{flushleft}
${ \begin{array}{ccccccccccccc}
   & &  & &  & &  P^{(3A)}
   & &  & &  & &  \cr
   & & & & & \nearrow & & \searrow & & & & & \cr
   P^{(0)} & \rightarrow & P^{(1)} & \Rightarrow & P^{(2)} & & & &
   P^{(4)}& \Rightarrow & P^{(5)} & \rightarrow & P^{(6)} \cr

   & & & & & \searrow & & \nearrow & & & & & \cr
\vspace{5 mm}
   & & & & & & P^{(3B)} & & & & & & \cr

   \end{array} }$\\
\end{flushleft}
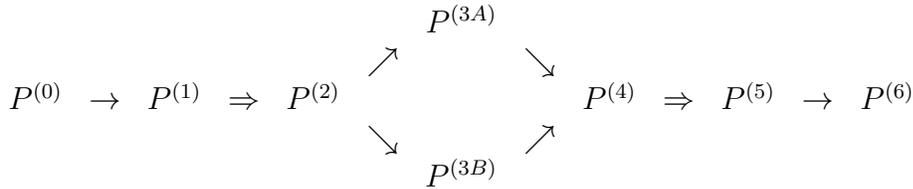
FIG.1. Instanton paths connecting critical points. The thick arrows show
there exist instanton effects. \\
\vspace{10mm}
\\
Now, the gradient flow equation is consist of simultaneous six differential
equations. It will be difficult to find the general solution of the
equation. However, we can find easily a pair of instanton solutions
connecting adjacent two critical points. In the next section we discuss
the instanton solutions and Hessian matrices. Transition amplitudes between
approximate vacuums are also computed.
\section{Instanton Solutions, Hessian Matrices and Transiton Amplitudes }
\setcounter{equation}{0}

{\bf A. $ P^{(0)} \rightarrow  P^{(1)} $} \\

In this subsection we examine instanton solutions connecting
$P^{(0)}$ and $P^{(1)}$. We find a pair of solutions satisfying (2.7).
The one is
\be
a \equiv c \equiv 0, b \equiv \pi,x=z=0,
y=\cos ^{-1}\tanh (c_{3}+c_{4})(t+\alpha),
\ee
where $y$ is the general solution of
$\displaystyle{dy \over dt}=-(c_{3}+c_{4})\sin y.$
The symbol $\equiv$ means that we can safely make the substitution
in $g^{ij}$. On the other hand, $x=z=0$ corresponds to singular points
in $g^{ij}$, but it is meaningful in the form $g^{ij} \partial _{j}h$.
The other is obtained by replacing $x=z=0$ by $x=z=\pi$ in (5.1).
Corresponding instanton paths are
\bea
  \left( \begin{array}{cccc}
        -1&  & & \\
          &-1& & \\
          & & \k y& \pm \s y \\
          & & \mp \s y & \k y \\
         \end{array} \right).
\eea
Since $\sin b =0$, the $6\times 6$ matrix $g^{ij}$ is reduced to a direct
sum of a $3\times 3$ matrix, a $2\times 2$ matrix and a $1\times 1$ matrix.
To obtain $H_{i}^{j}=H_{ik}g^{kj}$, we only have to calculate relevant
elements of $H_{ij}$. We note them in Appendix C.
The Hessian matrix $H_{i}^{j}$ for the first path is found to be \\
\bea
(H^{j}_{i}) =\\
\begin{tiny}
\bordermatrix{
\vspace{5 mm} & a & b & c & x & & z & y \cr
\vspace{5 mm}
a & c_{1} + \frac{c_{3}}2 - \frac{c_{3}+c_{4}}2 \cos y
  & \frac{c_{3}+c_{4}}2 (1-  \cos y)
  & \frac{c_{4}}2 & & & & \cr
\vspace{5 mm}
b & \frac{c_{2}+c_{3}}2 & c_{1} - c_{3}
  & \frac{c_{2}+c_{3}}2 & & & \bigzerou & \cr
\vspace{5 mm}
c & \frac{c_{2}+c_{4}}2 &  \frac{c_{3}+c_{4}}2 (-1+  \cos y)
  & c_{1} + \frac{c_{2}+c_{3}}2 - \frac{c_{3}+c_{4}}2 \cos y & & & & \cr
\vspace{5 mm}
x & & & & c_{2} - \frac{c_{3}+c_{4}}2 \cos y & & \frac{c_{3}-c_{4}}2 & \cr
\vspace{5 mm}
z & & \bigzerou & & \frac{c_{3}-c_{4}}2 & & c_{2}
  - \frac{c_{3}+c_{4}}2 \cos y & \cr
\vspace{5 mm}
y & & & & & & & -(c_{3}+c_{4}) \cos y \cr
  }
\end{tiny}  \nonumber
\eea
\be
\ee
where the indices $a, \cdots ,y$ represent quantum fluctuations
$\xi _{a}, \cdots , \xi _{y}$ around the classical solution (5.1).
Taking some linear combinations of the fluctuation coordinates,
the Hessian matrices $H_{i}^{j}$ can be diagonalized. Eigenvalues
of (5.3) are
\bea
\lambda _ {\tilde{a}}&=& c_{1} + \frac{c_{3}-c_{4}}2 - \frac{c_{3}+c_{4}}2 \cos
y ,
\quad
\lambda _{\tilde{c}}=c_{1}+\frac{c_{2}}2+
\frac{c_{3}+c_{4}}2 (1-  \cos y),\nonumber \\
\lambda _{ {x \pm z} \over \sqrt{2}}
&=&c_{2} \pm \frac{c_{3}-c_{4}}2 - \frac{c_{3}+c_{4}}2 \cos y ,\quad
\lambda _ {\tilde{b}}= c_{1}-c_{3}, \quad
\lambda _{y} =  -(c_{3}+c_{4})\cos y,
\eea
where $\tilde{a},\tilde{b}$ and $\tilde{c}$ are some linear combinations
of $a,b$ and $c$. The Hessian matrix for the second path is obtained by
changing signs of $H_{a}^{b},H_{b}^{a},H_{b}^{c}$ and $H_{c}^{b}$
in (5.3). Thus, the eigenvalues of $H_{i}^{j}$ are common to the two
instanton paths. Accordingly, bosonic transition amplitudes of the two
instanton solutions are equal. In the following subsections, we will see
this situation holds for any pair of instanton paths. We only have to
concentrate on fermionic contributions to find whether the two transition
amplitudes cancel each other or not. From (5.4) we see any eigenvalue
is positive at $t=-\infty$, and no fermionic mode is exited. So, the
approximate vacuum is a bosonic one
\be
|0>=\pi ^{-{3 \over 2}} \prod _{i=1}^{6} \lambda _{i}^{1 \over 4}
     \displaystyle{e}^{-{1 \over 2}\lambda _{i}\xi _{i}^{2}}|0>_{F},
\ee
where  $\lambda _{i}$ are eigenvalues of the Hessian matrices at
this point.  We represent by $|l>$ the approximate vacuum at
each critical point $P^{(l)}$. At $t=\infty$ the $y$ mode has the only \\
negative
eigenvalue. This implies that the fermionic mode corresponding to $y$ is
exited at $t=\infty$ and the following approximate vacuum
$|1>=\hat{\psi}_{y}^{*}|0>\sim dy|0>$ is induced. The second path
induces the following approximate vacuum $\hat{\psi}_{y^{\prime}}^{*}|0>
\sim dy^{\prime}|0>$. In this process $y$ is an odd
coordinate (See Appendix D.) and we have $dy^{\prime}=-dy$. The parity
of the state $|0>$ does not change and we see
\be
d_{h} |0>\sim (dy+dy^{\prime})|0>=0.
\ee
The operator $d_{h}$ increases the fermion number by 1. Since there is
no state with fermion number -1, $|0>$ can not be written as $|0>=d_{h}
|-1>$. Hence, $|0>$ is a true vacuum localized around $P^{(0)}$.
\vspace{10mm}
\\
{\bf B. $P^{(5)} \rightarrow P^{(6)}$}

In this subsection we examine instanton solutions connecting $P^{(5)}$
and $P^{(6)}$. Owing to the Poincar$\acute{e}$ duality[9], we can
easily perform analysis from the knowledge of the previous subsection.
We find a pair of solutions:
\be
a\equiv c \equiv b \equiv 0, y=\cos ^{-1}\tanh((c_{3}
+c_{4})t+\alpha),x=z=0 (x=z=\pi).
\ee
These solutions are represented by the following paths
\bea
  \left( \begin{array}{cccc}
         1&  & & \\
          & 1& & \\
          & & \k y& \pm \s y \\
          & & \mp \s y & \k y \\
         \end{array} \right).
\eea
{}From (C.1) we see $H_{ij}$ for these instanton solutions are obtained from
that of $P^{(0)} \rightarrow P^{(1)}$  by changing the signs except for
linear terms of $\cos y$. From (4.3) we see the metric $g^{ij}$ for
$P^{(5)} \rightarrow P^{(6)}$ is obtained from that of
$P^{(0)} \rightarrow P^{(1)}$  by changing the signs of linear terms of
$\cos y$. Accordingly, $H_{i}^{j}$ for $P^{(5)} \rightarrow P^{(6)}$ is
the opposite sign of (5.3) except for linear terms of $\cos y$. Thus,
eigenvalues of $H_{i}^{j}$ are
\bea
\lambda _ {\tilde{a}}&=& -c_{1} + \frac{c_{4}-c_{3}}2 - \frac{c_{3}+c_{4}}2
\cos y ,
\quad
\lambda _{\tilde{c}}=-c_{1}-\frac{c_{2}}2-
\frac{c_{3}+c_{4}}2 (1+  \cos y),\nonumber \\
\lambda _{ {x \pm z} \over \sqrt{2}}
&=&-c_{2} \mp \frac{c_{3}-c_{4}}2 - \frac{c_{3}+c_{4}}2 \cos y ,\quad
\lambda _ {\tilde{b}}= c_{3}-c_{1}, \quad
\lambda _{y} =  -(c_{3}+c_{4})\cos y.
\eea
At $t=-\infty$ any mode except for $y$ has a negative eigenvalue.
Therefore, at $t=-\infty$ the following fermionic state is an approximate
vacuum $|5>$ around $P^{(5)}$:
\bea
|5>\equiv |\tilde{a},\tilde{b},\tilde{c},{{x + z} \over \sqrt{2}},
{{x-z} \over \sqrt{2}}>\sim d\tilde{a}\wedge d\tilde{b}\wedge d\tilde{c}
\wedge d{{x + z} \over \sqrt{2}} \wedge d{{x-z} \over \sqrt{2}}|0> \nonumber \\
\sim  da\wedge db\wedge dc\wedge dx  \wedge dz |0>.
\eea
At $t=\infty$ the mode $y$ has a negative eigenvalue. This means a fermionic
mode corresponding to $y$ is excited. Thus, the first instanton path induces an
approximate vacuum $|6>$ around $P^{(6)}$:
\bea
|6>\equiv |\tilde{a},\tilde{b},\tilde{c},{{x + z} \over \sqrt{2}},
{{x-z} \over \sqrt{2}},y>
\sim  da\wedge db\wedge dc\wedge dx  \wedge dz  \wedge dy|0>.
\eea
In this process the coordinates
$a,c,x,z$ are even and $b,y$ are odd (See Appendix D.). Accordingly,
we see $|5>$ is odd and $|6>$ is even. Thus, from the general consideration in
section 3 we conclude
\be
<6|d_{h} |5>=0.
\ee
The number of exited fermions is at most 6, so we have $d_{h}|6>=0$.
Hence, $|6>$ is a true vacuum.
\vspace{10mm}
\\
{\bf C. $P^{(2)} \rightarrow P^{(3A)}$}

In this subsection we discuss instanton solutions connecting $P^{(2)}$
and $P^{(3A)}$. We find a pair of solutions:
\be
a\equiv c  \equiv 0, b \equiv \pi,
y=\cos ^{-1}\tanh(-((c_{3}
-c_{4})t+\alpha)),
x=0,z=\pi (x=\pi,z=0),
\ee
where $y$ is the general solution of
${dy \over dt}=(c_{3}-c_{4})\sin y$.
These solutions are represented as
\bea
  \left( \begin{array}{cccc}
        -1&  & & \\
          & 1& & \\
          & & -\k y& \mp \s y \\
          & & \mp \s y & \k y \\
         \end{array} \right).
\eea
Since $\sin b =0$, we can calculate $H_{i}^{j}$ as before.
For the first solution we find
\bea
(H^{j}_{i}) =\\
\begin{tiny}
\bordermatrix{
\vspace{5 mm} & a & b & c & x & & z & y \cr
\vspace{5 mm}
a & c_{1} - \frac{c_{2}}2 + \frac{c_{3}}2 ( \cos y -1)
  & \frac{c_{3}}2 (1-  \cos y)
  & \frac{c_{2}}2 & & & & \cr
\vspace{5 mm}
b & -\frac{c_{2}+c_{3}}2 & c_{1} + c_{3}
  & \frac{c_{2}+c_{3}}2 & & & \bigzerou & \cr
\vspace{5 mm}
c & \frac{c_{2}}2 &  \frac{c_{3}}2 (\cos y -1)
  & c_{1} - \frac{c_{2}}2 + \frac{c_{3}}2 (\cos y -1) & & & & \cr
\vspace{5 mm}
x & & & & -c_{2} - \frac{c_{4}}2 \cos y & & -c_{3}-\frac{c_{4}}2 & \cr
\vspace{5 mm}
z & & \bigzerou & & -c_{3}-\frac{c_{4}}2 & &- c_{2}
  - \frac{c_{4}}2 \cos y & \cr
\vspace{5 mm}
y & & & & & & & (c_{3}-c_{4}) \cos y \cr
  }
\end{tiny}  \nonumber ,
\eea
\be
\ee
For the second solution, the Hessiam matrix $H_{i}^{j}$ is obtained
by changing sings of $H_{a}^{b},H_{b}^{a},H_{b}^{c}$ and $H_{c}^{b}$ in
(5.15). Therefore, eigenvalues are common to the two paths. They are
\bea
\lambda _ {\tilde{a}}&=& c_{1} + \frac{c_{3}}2 (\cos y-1) ,\quad
\lambda _{\tilde{c}}=c_{1}+\frac{c_{3}}2 (1+  \cos y),\nonumber \\
\lambda _{{x \pm  z} \over \sqrt{2}}
&=&-c_{2} \pm c_{3} \pm \frac{c_{4}}2 (1-\cos y) ,\quad
\lambda _ {\tilde{b}}= c_{1}-c_{2}, \quad
\lambda _{y} =  (c_{3}-c_{4})\cos y.
\eea
At $t=-\infty$ fermionic modes corresponding to ${{x \pm  z} \over \sqrt{2}}$
are exited. At $t=\infty$ fermionic mode corresponding to $y$ is also
exited. Thus, the instantons induce
\be
|{{x+  z} \over \sqrt{2}},{{x-  z} \over \sqrt{2}}> \longrightarrow
|{{x+  z} \over \sqrt{2}},{{x-  z} \over \sqrt{2}},y>
\ee
Now, the coordinates $x,z,a,c$ are even and $b,y$ are odd(See Appendix D.).
So, the original state is odd and the new one is even. From the
general consideration we conclude there is no instanton effect in the process
$P^{(2)} \rightarrow P^{(3A)}$.
\vspace{10mm}
\\
{\bf D. $P^{(3B)} \rightarrow P^{(4)}$}

{}From the duality of $P^{(2)} \rightarrow P^{(3A)}$ we find two solutions
connecting $P^{(3B)}$ and $P^{(4)}$:
\be
a \equiv c \equiv b \equiv 0,y=\cos ^{-1}(-\tanh ((c_{3}-c_{4})t+\alpha))
,
x=0,z=\pi (x=\pi, z=0).
\ee
These solutions are represented by the paths
\bea
  \left( \begin{array}{cccc}
         1&  & & \\
          & -1& & \\
          & &- \k y& \mp \s y \\
          & & \mp \s y & \k y \\
         \end{array} \right).
\eea
In this process $x,z,a,c$ are even coordinates and $b,y$ are odd ones(See
Appendix D.). Eigenvalues of the corresponding Hessian matrix are
\bea
\lambda_{a+c \over \sqrt{2}}&=& -c_{1} + \frac{c_{3}}2( \cos y +1) ,\quad
\lambda _{a-c+b\over \sqrt{2}}
=c_{1}+\frac{c_{3}}2 (\cos y- 1 ),\nonumber \\
\lambda _{ {x \pm z} \over \sqrt{2}}
&=&c_{2} \mp c_{3} - \frac{c_{4}}2(\cos y \pm 1) ,\quad
\lambda _ {\tilde{b}}= -c_{1}+c_{2}, \quad
\lambda _{y} =  (c_{3}-c_{4})\cos y,
\eea
where $\tilde{b}$ denotes $b$ at $t=-\infty $ and $b+{c_{3} \over
{c_{2}+c_{3}}}(a-c)$ at $t=\infty$. From (5.20) we find the instantons
induce the following transition between approximate vacuums:
\be
|{a+c \over \sqrt{2}},b,{a-c+b\over \sqrt{2}}>
\rightarrow
|{a+c\over \sqrt{2}},b+{c_{3} \over{c_{2}+c_{3}}}(a-c)
,{a-c+b\over \sqrt{2}},y>.
\ee
The original and the new states are essentially an odd one
$da \wedge db \wedge dc |0>$ and an even one
$da \wedge db \wedge dc \wedge dy |0>$ respectively.
Thus, there is no instanton effect in the process $P^{(3B)}$ and $P^{(4)}$.
\vspace{10mm}
\\
{\bf E. $P^{(2)} \rightarrow P^{(3B)}$}

In this subsection we discuss instanton solutions connecting $P^{(2)}$
and $P^{(3B)}$. Setting $b \equiv 0, y=0,x \equiv z \equiv
 \displaystyle{\pi / 2}$,
the gradient flow equation (2.7) reduces to
\bea
{d(a-c) \over dt}&=&-(c_{1}-c_{2})\sin (a-c), \nonumber \\
{d(a+c) \over dt}&=&-(c_{1}+c_{2})\sin (a+c).
\eea
Setting $a+c \equiv 0$, we get the following solutions:
\be
b \equiv 0, y=0,x \equiv z \equiv \displaystyle{\pi / 2},a+c \equiv 0 ,
a-c = \pm \cos ^{-1}\tanh ((c_{1}-c_{2})t+\alpha) .
\ee
These are represented by the paths
\bea
  \left( \begin{array}{cccc}
         \k (a-c)&\pm \s (a-c)  & & \\
         \pm \s (a-c) & -\k (a-c)& & \\
          & & -1&  \\
          & &  & 1 \\
         \end{array} \right).
\eea
Since $\sin b \equiv \cos x \equiv \cos z \equiv 0$, the metric $g^{ij}$
reduces to a direct sum of two $2\times 2$ matrices and two $1 \times 1$
matrices. Relevant $H_{ij}$ to obtain $H_{i}^{j}$ are given in (C.2).
We get the following Hessian matrix for the both solutions. Non-zero
matrix elements are
\bea
(H_{i}^{j})=\bordermatrix{
\vspace{5mm} & a & c& \cr
 a &c_{1}({\s ^{2} a \over 2} -\k ^{2} a)+c_{2}({\k ^{2} a \over 2} -\s ^{2} a)
  &-{c_{1} \over 2}\s ^{2}a-{c_{2} \over 2}\k ^{2}a & \cr
\vspace{5mm}
c &-{c_{1} \over 2}\s ^{2}a-{c_{2} \over 2}\k ^{2}a
 &c_{1}({\s ^{2} a \over 2} -\k ^{2} a)+c_{2}({\k ^{2} a \over 2} -\s ^{2} a)
 & \cr
} \nonumber ,
\eea
\bea
(H_{i}^{j})=\bordermatrix{
\vspace{5mm} & x & z & \cr
 x &{c_{1} \over 2}\s ^{2} a  +{c_{2} \over 2} \k ^{2} a +c_{3}-{c_{4} \over 2}
  &{c_{1} \over 2}\s ^{2} a  +{c_{2} \over 2} \k ^{2} a +{c_{4} \over 2}
 & \cr
\vspace{5mm}
z &{c_{1} \over 2}\s ^{2} a  +{c_{2} \over 2} \k ^{2} a +{c_{4} \over 2}
 &{c_{1} \over 2}\s ^{2} a  +{c_{2} \over 2} \k ^{2} a +c_{3}-{c_{4} \over 2}
 & \cr
} \nonumber ,
\eea
\bea
H_{b}^{b}=-c_{1}\k ^{2} a -c_{2}\s ^{2} a +c_{3}, \qquad
H_{y}^{y}=c_{1}\s ^{2} a +c_{2}\k ^{2} a -c_{4}.
\eea
Eigenvalues of this matrix are
\bea
\lambda _{{a+c} \over \sqrt{2}}&=&-c_{1}\k ^{2} a-c_{2}\s ^{2} a ,\qquad
\lambda _{{a-c} \over \sqrt{2}}=(c_{2}-c_{1})\k (a-c) , \nonumber \\
\lambda _{{x+z} \over \sqrt{2}}&=&c_{1}\s ^{2} a+c_{2}\k ^{2} a,
\qquad
\lambda _{ {x- z} \over \sqrt{2}}
=c_{3}-c_{4} ,\nonumber \\
\lambda _ {b}&=&-c_{1}\k ^{2} a-c_{2}\s ^{2} a +c_{3}, \qquad
\lambda _{y} = c_{1}\s ^{2} a+c_{2}\k ^{2} a-c_{4}.
\eea
{}From (5.26) we see the instanton solutions cause the following transition:
\be
| {{a+c} \over \sqrt{2}},b> \rightarrow
| {{a+c} \over \sqrt{2}},b,{{a-c} \over \sqrt{2}}>.
\ee
In this process $a,c $ are odd coordinates and $x,y,z,b$ are even ones and
the original state is odd and the new state is even. Hence, the two instanton
effects cancel each other.
\vspace{10mm}
\\
{\bf F. $P^{(3A)} \rightarrow P^{(4)}$}

Setting $b \equiv \pi, y= \pi$ and changing $a \rightarrow -a,
c \rightarrow -c $ in (5.23) we obtain two solutions connecting $P^{(3A)}$
and $P^{(4)}$. Corresponding paths are
\bea
  \left( \begin{array}{cccc}
         -\k (a-c)&\mp \s (a-c)  & & \\
         \mp \s (a-c) & \k (a-c)& & \\
          & & 1&  \\
          & &  & -1 \\
         \end{array} \right).
\eea
The Hessian matrix $H_{i}^{j}$ and its eigenvalues are given by the
opposite sign of (5.25) and (5.26) respectively. The following transition
is induced:
\be
| {{x+z} \over \sqrt{2}},{{x-z} \over \sqrt{2}},y> \rightarrow
|{{x+z} \over \sqrt{2}},{{x-z} \over \sqrt{2}} ,y,{{a-c} \over \sqrt{2}}>.
\ee
The parities of the coordinates are the same as in the previous subsection.
These two states have opposite parities. Hence,there is no instanton
effect.
\vspace{10mm}
\\
{\bf G. $P^{(1)} \rightarrow P^{(2)}$}

As we have seen there is no instanton effect in the previous six
processes. In the following two processes we will see there exist
instanton effects.

We find two instanton solutions connecting $P^{(1)}$ and $P^{(2)}$:
\be
a \equiv -c \equiv \pm {\pi \over 2},x \equiv z \equiv {\pi \over 2},
y=0,b=\k ^{-1}(-\tanh((c_{2}-c_{3})t+\alpha)).
\ee
Corresponding paths are
\bea
  \left( \begin{array}{cccc}
         -1&  & & \\
           & \k b &\pm \s b & \\
          & \pm \s b& -\k b&  \\
          & &  & 1 \\
         \end{array} \right).
\eea
In this process $a,c$ are even  coordinates and $x,z,y,b$ are odd ones.
Since  $y=0$ corresponding to a singular metric, we set $y=\epsilon $.
Then we have
\bea
(g^{ij})=  \nonumber  \\
 \bordermatrix{
\vspace{5 mm} & a & b & c & x & z & y \cr
\vspace{5 mm}
a & \displaystyle{\frac1{\epsilon^{2}} + \frac13} & 0
  & \pmatrix{ \displaystyle{\frac1{\epsilon^{2}} - \frac16}} \cos b & 0
  & \pmatrix{ - \displaystyle{\frac1{\epsilon^{2}} + \frac16}} \sin b & 0 \cr
\vspace{5 mm}
b & 0 & 1 & 0 & 0 & 0 & 0 \cr
\vspace{5 mm}
c & \pmatrix{ \displaystyle{\frac1{\epsilon^{2}} - \frac16}} \cos b & 0
  & \displaystyle{\frac1{\epsilon^{2}} + \frac13}
  & \pmatrix{ - \displaystyle{\frac1{\epsilon^{2}} + \frac16}} \sin b
  & 0 & 0 \cr
\vspace{5 mm}
x & 0 & 0 & \pmatrix{ - \displaystyle{\frac1{\epsilon^{2}} + \frac16}} \sin b
  & \displaystyle{\frac1{\epsilon^{2}} + \frac13}
  & \pmatrix{ - \displaystyle{\frac1{\epsilon^{2}} + \frac16}} & 0 \cr
\vspace{5 mm}
z & \pmatrix{ - \displaystyle{\frac1{\epsilon^{2}} + \frac16}} \sin b & 0 & 0
  & \pmatrix{ - \displaystyle{\frac1{\epsilon^{2}} + \frac16}} \cos b
  & \displaystyle{\frac1{\epsilon^{2}} + \frac13} & 0 \cr
\vspace{5 mm}
y & 0 & 0 & 0 & 0 & 0 & 1 \cr
  } .
\eea
\be
\ee
This expression shows that the modes $b$ and $y$ do not combine
with another mode. Since $g^{ij}$ and $\Gamma _{ij}^{k}$ are
independent of $a$ and $c$, $H_{ij}$ and $H_{i}^{j}$ are common to
the two solutions. From (5.32) and (C.3) we get
\bea
 (H^{j}_{i}) = \nonumber \\
\bordermatrix{
\vspace{5 mm} & a & c & & x & z & b &  y \cr
\vspace{5 mm}
a & \displaystyle{\frac{c_{1}}2}-c_{2} \cos b
  & - \displaystyle{\frac{c_{1}}2} \cos b
  & & \displaystyle{\frac{c_{2}+c_{3}}2} \sin b
  & \displaystyle{\frac{c_{1}}2} \sin b &  & \cr
\vspace{5 mm}
c & - \displaystyle{\frac{c_{1}}2} \cos b
  & \displaystyle{\frac{c_{1}}2}-c_{2} \cos b
  & & \displaystyle{\frac{c_{1}}2} \sin b
  & \displaystyle{\frac{c_{2}+c_{3}}2} \sin b & & \bigzerou  \cr
\vspace{5 mm}
x & \displaystyle{\frac{c_{2}+c_{3}}2} \sin b
  & \displaystyle{\frac{c_{1}}2} \sin b
  & & \displaystyle{\frac{c_{1}}2}+c_{3} \cos b
  & \displaystyle{\frac{c_{1}}2} \cos b &  & \cr
\vspace{5 mm}
z & \displaystyle{\frac{c_{1}}2} \sin b
  & \displaystyle{\frac{c_{2}+c_{3}}2} \sin b
  & & \displaystyle{\frac{c_{1}}2} \cos b
  & \displaystyle{\frac{c_{1}}2}+c_{3} \cos b & &  \cr
\vspace{5 mm}
b & & & & & & (c_{3}-c_{2}) \cos b &  0 \cr
\vspace{5 mm}
y & & & \bigzerou& & & 0  & c_{1}-c_{4} \cr
  }.
\eea

\be
\ee
Eigenvalues of (5.33) are
\bea
2\lambda _{1}&=&2c_{1}-c_{2}-c_{3}+(c_{3}-c_{2})\k b, \qquad
2\lambda _{2}= c_{2}+c_{3}+(c_{3}-c_{2})\k b,  \nonumber \\
2\lambda _{3}&=& 2c_{1}+c_{2}+c_{3}+(c_{3}-c_{2})\k b, \qquad
2\lambda _{4}= -c_{2}-c_{3}+(c_{3}-c_{2})\k b, \nonumber \\
\lambda _{b}&=&(c_{3}-c_{2})\k b, \qquad
\lambda _{y}=c_{1}-c_{4}.
\eea
Linear combinations of modes $a,c,x$ and $c$ correspond to
the first four eigenvalues. Only the fourth eigenvalue $\lambda _{4}$
is negative. At $t=-\infty $ the negative eigenvalue is $-c_{3}$ and
the corresponding mode is ${x+z \over \sqrt{2}}$, which is an odd coordinate.
At $t=\infty $ the negative eigenvalue has changed to $-c_{2}$ and the
corresponding mode is ${a-c \over \sqrt{2}}$, which is an even coordinate.
The excited eigenmode  has changed from the odd one to the even one.
This is a new feature that can not be seen in the previous six
processes. Thus, the instanton solutions cause transition between odd states:
\be
| {x+z \over \sqrt{2}}> \rightarrow
| {a-c \over \sqrt{2}},b>.
\ee
{}From the general argument in Sect.3 the two instanton effects do not cancel
each other.
\vspace{10mm}
\\
{\bf H. $P^{(4)} \rightarrow P^{(5)}$}

Replacing $y=0$ by $y=\pi$ in (5.30), we obtain
two instanton solutions connecting $P^{(4)}$ and $P^{(5)}$.
These solutions are represented as
\bea
  \left( \begin{array}{cccc}
         1&  & & \\
           & -\k b &\mp \s b & \\
          & \pm \s b& \k b&  \\
          & &  & -1 \\
         \end{array} \right).
\eea
Corresponding Hessian matrix $H_{i}^{j}$ is obtained by the following
replacements $c_{1} \rightarrow -c_{1}, c_{4} \rightarrow -c_{4}$ in the
diagonal part of $H_{i}^{j}$ (5.33).
Eigenvalues of $H_{i}^{j}$ are
\bea
2\lambda _{1}&=&c_{2}+c_{3}+(c_{3}-c_{2})\k b,\qquad
2\lambda _{2}= -2c_{1}+c_{2}+c_{3}+(c_{3}-c_{2})\k b,  \nonumber \\
2\lambda _{3}&=& -c_{2}-c_{3}+(c_{3}-c_{2})\k b,\qquad
2\lambda _{4}= 2c_{1}-c_{2}-c_{3}+(c_{3}-c_{2})\k b, \nonumber \\
\lambda _{b}&=&(c_{3}-c_{2})\k b, \qquad
\lambda _{y}=c_{4}-c_{1}.
\eea
At $t=-\infty $ there are three negative eigenvalues $-c_{1}+c_{2},-c_{3}$
and $-c_{1}-c_{3}$ except for $c_{4}-c_{1}$.
Corresponding modes are ${a-c \over \sqrt{2}}$, ${x-z \over \sqrt{2}}$
and ${x+z \over \sqrt{2}}$ respectively. The three eigenvalues gradually
change to  $-c_{1}+c_{3},-c_{2}$ and
 $-c_{1}-c_{2}$ respectively at $t=\infty $ . Corresponding modes are
${x-z \over \sqrt{2}}$, ${a-c \over \sqrt{2}}$
and ${a+c \over \sqrt{2}}$ respectively. We have now the following
transition of states:
\be
|{a-c \over \sqrt{2}},{x-z \over \sqrt{2}}, {x+z \over \sqrt{2}},y>
 \rightarrow
|{x-z \over \sqrt{2}},{a-c \over \sqrt{2}} {a+c \over \sqrt{2}},y,b>.
\ee
The parities of the coordinates are the same as in the previous
subsection and the parities of both states are odd.
Hence,there exist instanton effects.
\section{Summary and Discussion}

We have investigated instanton solutions connecting the approximate
vacuums on $SO(4)$. For any pair of adjacent approximate vacuums we have found
two instanton paths and have discussed the transition amplitude.
We have found in the two processes instanton effects
exist. We have given the criterion for the presence of instanton
effects; if the parities of the two approximate vacuums are identical,
there exist instanton effects. If they are opposite, the two instanton
effects cancel each other. Two approximate vacuums coupled by instanton
effects are not true vacuums. The approximate vacuums around
$P^{(1)},P^{(2)},P^{(4)}$ and $P^{(5)}$ are not true vacuums. We have left
the four true vacuums around $P^{(0)},P^{(3A)},P^{(3B)}$ and $P^{(6)}$.
This result is in agreement with the de Rham cohomology of $SO(4)$.

In the $SO(3)$ case, two time dependent harmonic oscillator modes
accompany with one instanton mode[6]. One definite Euler angle
corresponds to the instanton mode in every process.
In the $SO(4)$ case,a particular coordinate corresponds to the instanton mode
in each process. The other five coordinates except for the
instanton mode play the role of
time dependent harmonic oscillators.

We have found a pair of instanton solutions connecting two critical
points with Morse index differs by one. We do not have discussed the
general solution of the gradient flow equation (2.7). However, I believe
there are not any other solutions connecting the critical points.
To make this point clear, further investigation will be needed.
\vspace{20mm}
\\
{\bf Acknowledgments}

The author thanks for Dr.K.Kuribayashi for useful discussion.
He is indebted to Dr.Y.Yasui for valuable discussion.

\setcounter{equation}{0}
\appendix
\section*{Appendix A: Left Invariant Vector Fields and an Invariant Metric}
{}From the expression of the group element $A$ (4.2) left invariant
vector fields $e_{ij}$ corresponding to $E_{ij}$ can be read as
\begin{eqnarray*}
e_{12} &=&{\partial \over \partial a}, \qquad \qquad
e_{23}=-{\s a \k x \over \s x }{\partial \over \partial a}
       +{\s a  \over \s x }{\partial \over \partial b}
       +\k a {\partial \over \partial x}, \nonumber \\
e_{34} &=&-{\s b \s x \k z \over \s y \s z }{\partial \over \partial b}
+{\s b \s x  \over \s y \s z }{\partial \over \partial c}
-{\s x \k y  \over \s y }{\partial \over \partial x}
+\k x {\partial \over \partial y}
+{\k b \s x  \over \s y }{\partial \over \partial z},\nonumber \\
e_{13} &=&-{\k a \k x \over \s x }{\partial \over \partial a}
+{\k a  \over \s x }{\partial \over \partial b}
-\s a{\partial \over \partial x}, \nonumber \\ \vspace{5mm}
e_{14} &=& {\k a \k y \over \s x \s y }{\partial \over \partial a}
       -({\k a \k x \k y \over \s x \s y} +
{{\k a \k b \k z-\s a \s b \k x \k z} \over \s y \s z})
{\partial \over \partial b} \nonumber \\
&+& {{\k a \k b -\s a \s b \k x} \over \s y \s z}
{\partial \over \partial c}
+\s a \k x {\k y \over \s y}{\partial \over \partial x}
+\s a \s x{\partial \over \partial y} \nonumber \\
&-& {{\k a \s b +\s a \k b \k x } \over \s y }
{\partial \over \partial z}, \nonumber \\
e_{24} &=& {\s a \k y \over \s x \s y }{\partial \over \partial a}
-\k a \k x{\k y \over \s y}{\partial \over \partial x}
-\k a \s x{\partial \over \partial y} \nonumber \\
&+&{{\k a \k b \k x -\s a \s b } \over \s y }
{\partial \over \partial z}.  \hspace{77mm} \rm{(A.1)}
\end{eqnarray*}
An $SO(4)$ invariant metric is now given by
\begin{eqnarray*}
\hspace{30mm}
g^{ij}={\displaystyle \sum_{(\alpha \beta)}} e_{\alpha \beta}^{i}
e_{\alpha \beta}^{j}.  \hspace{89mm} {\rm (A.2)}
\end{eqnarray*}
\setcounter{equation}{0}
\section*{Appendix B: Christoffel Symbols}
In this appendix, we summarize 38 non-zero Christoffel symbols under
the metric (4.4). We use them in computing the Hessian $H_{ij}$.
\begin{eqnarray*}
\Gamma_{x}^{z}{}_{y}&=&-\Gamma_{xz}^{y}=-\frac{1}{2} \k b \s y ,\qquad
\Gamma_{xy}^{x}=\Gamma_{yz}^{z}=\Gamma_{ya}^{a}=\Gamma_{yc}^{c}={\k y \over 2\s
y} , \nonumber \\
\Gamma_{xy}^{b} &=& {\s b\k z \over 2\s y \s z} ,\quad
\Gamma_{xy}^{b}={\s b \over 2\s y \s z} ,\quad
\Gamma_{xz}^{b}=\frac{1}{2} \s b \k y ,\quad
\Gamma_{xa}^{b}=\Gamma_{xb}^{a}=-{1 \over 2\s x} , \nonumber \\
\Gamma_{xa}^{a} &=& \Gamma_{xb}^{b}={\k x \over 2\s x} , \qquad
\Gamma_{xc}^{y}=\frac{1}{2} \s b \s y \s z,\qquad
\Gamma_{xc}^{b}=-\frac{1}{2} \k b \k y \s z +{\k x \k z \over 2\s x},\nonumber
\\
\Gamma_{xc}^{a} &=& {\k x \k z \over 2\s x} ,\qquad
\Gamma_{yz}^{x}=-{\k b \over 2\s y} ,  \qquad
\Gamma_{yz}^{a}=-{\s b \over 2\s x\s y} ,\qquad
\Gamma_{yz}^{b}={\s b \k x \over 2\s x\s y} ,\nonumber \\
\Gamma_{ya}^{z} &=&-{\s b \s x \over 2\s y} ,\qquad
\Gamma_{ya}^{b}=-{\k x \k y \over 2\s y} -{\k b \s y \k z \over 2\s y \s z} ,
\qquad
\Gamma_{ya}^{c}={\k b \s x \over 2\s y \s z} ,\nonumber \\
\Gamma_{yc}^{x} &=&-{\s b \s z \over 2\s y } ,\qquad
\Gamma_{yc}^{b}=-{\k y \k z \over 2\s y} -{\k b \k y \s z \over 2\s x \s y}
,\qquad
\Gamma_{yc}^{a}={\k b \s z \over 2\s x \s y } ,\nonumber \\
\Gamma_{za}^{y} &=& \frac{1}{2} \s b \s x \s y,\qquad
\Gamma_{xc}^{b}=-\frac{1}{2} \k b \s x \k y +{\k x \k z \over 2\s z},\qquad
\Gamma_{za}^{c}=-{\k x \over 2\s z } ,\nonumber \\
\Gamma_{zb}^{b} &=& \Gamma_{zc}^{c}={\k z \over 2\s z } ,\qquad
\Gamma_{zb}^{c}=\Gamma_{zc}^{b}=-{1 \over 2\s z } ,\qquad
\Gamma_{ab}^{x}=\frac{1}{2} \s x ,\nonumber \\
\Gamma_{ac}^{x} &=& \frac{1}{2} \s x \k z ,\qquad
\Gamma_{ac}^{y}=-\frac{1}{2} \k b \s x \s y \s z,\qquad
\Gamma_{ac}^{z}=\frac{1}{2} \k x \s z ,\nonumber \\
\Gamma_{ac}^{b} &=&-\frac{1}{2}\s b \s x \k y \s z , \qquad
\Gamma_{bc}^{z} =\frac{1}{2}\s z .
\hspace{54mm}  \rm{(B.1)}
\end{eqnarray*}
\setcounter{equation}{0}
\section*{Appendix C: Relevant Hessian Matrices}
In this appendix, we note relevant Hessian matrices $H_{ij}$.

The following $H_{ij}$ are necessary in computing $H_{i}^{j}$
for the processes $P^{(0)} \rightarrow P^{(1)},P^{(5)} \rightarrow
P^{(6)},P^{(2)} \rightarrow P^{(3A)}$ and
$P^{(3B)} \rightarrow P^{(4)}$:
\begin{eqnarray*}
H_{aa} &=&H_{cc}=-c_{1}\k b +c_{2}(\s x \s z \k y- \k b \k x \k z), \nonumber
\\
H_{ab} &=&-c_{1}\k b \k x +\frac{c_{2}}{2}( \k b \s ^{2}x \k z
+\s x \s z \k x \k y - 2\k b  \k z) \nonumber
\\ &+& \frac{c_{3}}{2}( \s ^{2}x \k z \k y +\k b \s x \s z \k x), \nonumber \\
H_{ac} &=&c_{1}(\s x \s z \k y- \k b \k x \k z) +\frac{c_{2}}{2}( \k b (\s
^{2}x \k ^{2}z+\k ^{2}x \s ^{2}z-2)
\nonumber \\
&+&2\s x \s z \k x\k z\k y)
-\frac{1}{2}(c_{3}\k x \k z+c_{4})\k b \s x\s z \s ^{2} y
\nonumber
\\ &-&\frac{c_{3}}{2}(\k y (\s ^{2}x \k ^{2}z+\k ^{2}x \s ^{2}z)
+\k b\s x \s z \k x \k z),\nonumber \\
H_{bb} &=&-c_{1}\k b-c_{2}\k b \k x \k z+c_{3}\k b \s x\s z,
\nonumber \\
H_{bc} &=&-c_{1}\k b \k z +\frac{c_{2}}{2}( \k b \k x\s ^{2}z
+\s x \s z \k z \k y - 2\k b  \k x), \nonumber \\
H_{xx} &=& H_{zz}=-\k x \k z(c_{2}\k b+c_{3}\k y),\nonumber \\
H_{xz} &=&-\k x \k z(c_{2}\k y+c_{3}\k b)
-\frac{1}{2}(c_{3}\k x \k z+c_{4})\k b \s ^{2} y,\nonumber \\
H_{yy} &=&-(c_{3}\k x \k z+c_{4})\k y .
\hspace{81mm} \rm{(C.1)}
\end{eqnarray*}
Note that we have not discarded the terms containing $\s x$ or $\s z$
which can survive combined with the metric $g^{ij}$ .

We use the following $H_{ij}$ in the processes $P^{(2)} \rightarrow P^{(3B)}$
and $P^{(3A)} \rightarrow P^{(4)}$:
\begin{eqnarray*}
H_{aa} &=& H_{cc}=c_{1}(\s ^{2}a\k y- \k ^{2}a \k b)
+c_{2}(\k ^{2}a \k y-\s ^{2}a\k b ), \nonumber \\
H_{ac} &=&c_{1}(\k ^{2}a \k y-\s ^{2}a\k b )+c_{2}(\s ^{2}a\k y -\k ^{2}a \k b)
\nonumber \\
&+&\frac{1}{2}\k b \s ^{2}y(c_{1}\s ^{2}a+c_{2}\k ^{2}a-c_{4}), \nonumber \\
H_{bb} &=&-c_{1}\k a \k a \k c+c_{2}\s a \k b \s c+c_{3}\k b, \nonumber \\
H_{xx} &=& H_{zz}=c_{1}\s ^{2}a \k y+c_{2}\k ^{2}a \k y+c_{3}\k b, \nonumber \\
H_{xz} &=& \frac{1}{2}\k b \s ^{2}y(-c_{1}\s ^{2}a-c_{2}\k ^{2}a+c_{4}),
\nonumber \\H_{yy} &=& c_{1}\s ^{2}a \k y+c_{2}\k ^{2}a \k y-c_{4}\k y,
\hspace{47mm} \rm{(C.2)}
\end{eqnarray*}
where we have used $c \equiv -a$.

In the processes $P^{(1)} \rightarrow P^{(2)}$ and $P^{(4)} \rightarrow
P^{(5)}$
the following $H_{ij}$ are necessary:
\begin{eqnarray*}
H_{aa} &=& H_{cc}=c_{1}\k y -c_{2}\k b, \nonumber \\
H_{ac} &=& c_{1}(\frac{1}{2}\s ^{2}y-1)\k b+c_{2}(1-\frac{1}{2}\s ^{2}b)\k y
+\frac{c_{3}}{2}\k y \s ^{2}b, \nonumber \\
H_{ax} &=&H_{cz}=\frac{c_{2}+c_{3}}{2}\s b, \nonumber \\
H_{az} &=& H_{cx}=c_{1}(1-\frac{1}{2}\s ^{2}y)\s b
+\frac{c_{3}-c_{2}}{2}\k y\k b\s b, \nonumber \\
H_{xx} &=& H_{zz}=c_{1}\k y +c_{3}\k b, \nonumber \\
H_{xz} &=& (1-\frac{1}{2}\s ^{2}y)(c_{1}\k b+c_{3}\k y)
+\frac{c_{2}}{2}\k y\s ^{2}b, \nonumber \\
H_{bb} &=& (c_{3}-c_{2})\k b, \qquad H_{yy}=(c_{1}-c_{4})\k y.
\hspace{46mm} \rm{(C.3)}
\end{eqnarray*}
\setcounter{equation}{0}
\section*{Appendix D: Parities  of Coordinates}
In this appendix, we identify parities of coordinates.

We discuss the process $P^{(0)} \rightarrow P^{(1)}$ as an example.
Around the first instanton solution (5.1) connecting
$P^{(0)}$ and $P^{(1)}$,the group element $A$ is of the form
up to second order of the fluctuation coordinates:
\vspace{5mm}
\begin{eqnarray*}
  \left( \begin{array}{cccc}
         -1&-a-b-c&0 &0 \\
        a+b+c& -1 &-x+z \k y_{0} & 0\\
         0 & z-x\k y_{0}& \k y_{0}& y\k y_{0} \\
         0 & 0& -y\k y_{0} & \k y_{0} \\
         \end{array} \right),\hspace{49mm} \rm{(D.1)}
\vspace{5mm}
\end{eqnarray*}
where $y_{0}$ denotes the classical solution and
$a,b, \ldots ,y$ are abbreviations of the fluctuation coordinates.
On the other hand, around the second solution of $P^{(0)} \rightarrow P^{(1)}$
$A$ is of the form:
\vspace{5mm}
\begin{eqnarray*}
  \left( \begin{array}{cccc}
         -1&-a+b-c&0 &0 \\
        a-b+c& -1 &-x+z \k y_{0} & 0\\
         0 & z-x\k y_{0}& \k y_{0}& -y\k y_{0} \\
         0 & 0& y\k y_{0} & \k y_{0} \\
         \end{array} \right).\hspace{47mm}\rm{(D.2)}
\vspace{5mm}
\end{eqnarray*}
Comparing (D.1) and (D.2), we find $a,c,x$ and $z$ are even coordinates
and $b$ and $y$ are odd ones in this process.

In the same way, we can identify parities of the coordinates in each process.
Results are summarized in TABLE 1.
\vspace{10mm}
\\
\vspace{10mm}
\begin{center}
TABLE 1: Parities of coordinates.\\
\vspace{5mm}
\begin{tabular}{|c|c|c|} \hline
Processes & even coordinates & odd coordinates \\ \hline
     &  &  \\
$ P^{(0)} \rightarrow P^{(1)}, P^{(5)} \rightarrow P^{(6)} $  &  &   \\
     & a,c,x,z & b,y \\
$ P^{(2)} \rightarrow P^{(3A)}, P^{(3B)} \rightarrow P^{(4)} $ &  &   \\
 & & \\ \hline
 & & \\
$ P^{(2)} \rightarrow P^{(3B)}, P^{(3A)} \rightarrow P^{(4)} $ & b,x,z,y & a,c
\\
 & & \\ \hline
 & & \\
$ P^{(1)} \rightarrow P^{(2)}, P^{(4)} \rightarrow P^{(5)} $& a,c& b,x,z,y \\
 & & \\ \hline
\end{tabular}
\\
\end{center}
\newpage
{\bf REFERENCES}  \\
$[1]$ See for a review,M.Kaku, Strings,Conformal Fields and Topology:
An Introduction\\(Springer-Verlag, New York,1991). \\
$[2]$ See, for example, E.Witten,Princeton University Report No.
IASSNS-94-96,1994.
\\
$[3]$ E.Witten,J.Diff.Geom.{\bf 17},661(1982). \\
$[4]$ B.Helffer and J.Sj\"{o}strand,Comm.in P.D.E.{\bf 10(3)},245(1985). \\
$[5]$ P.Salomonson and J.W.Van Holten, Nucl.Phys.B{\bf 196},509(1982). \\
$[6]$ T.Hirokane,M.Miyajima and Y.Yasui,J.Math.Phys.{\bf 34},2789(1993). \\
$[7]$ I.Yokota,Manifold and Morse Theory(Gendai Suugakusha,Kyoto,1989
(in Japanese)). \\
$[8]$ M.B\"{o}hm and G.Junker,J.Math.Phys.{\bf 28},1978(1987). \\
$[9]$ See, for example ,M.Kaku,Introduction to Superstrings Chap 11
(Springer-Verlag, \\New Yoyk,1988). \\
\end{document}